\documentclass[12pt]{article}
\usepackage{epsfig,amssymb}

\newcommand{\bq}{\begin{equation}}
\newcommand{\eq}{\end{equation}}
\newcommand{\bqa}{\begin{eqnarray}}
\newcommand{\eqa}{\end{eqnarray}}
\newcommand{\ben}{\begin{enumerate}}
\newcommand{\een}{\end{enumerate}}
\newcommand{\bc}{\begin{center}}
\newcommand{\ec}{\end{center}}
\newcommand{\bqb}{\begin{eqnarray*}}
\newcommand{\eqb}{\end{eqnarray*}}

\def\lsim{\lesssim}

\def\pr#1#2#3{ Phys. Rev. ${\bf{#1}}$ (#2) #3}
\def\prl#1#2#3{ Phys. Rev. Lett. ${\bf{#1}}$ (#2) #3}
\def\pl#1#2#3{ Phys. Lett. ${\bf{#1}}$ (#2) #3}
\def\prep#1#2#3{ Phys. Rep. ${\bf{#1}}$ (#2) #3}

\def\np#1#2#3{ Nucl. Phys. ${\bf{#1}}$ (#2) #3}
\def\zp#1#2#3{ Z. f. Phys. ${\bf{#1}}$ (#2) #3}
\def\epj#1#2#3{ Eur. Phys. J. ${\bf{#1}}$ (#2) #3}
\def\ijmp#1#2#3{ Int. J. Mod. Phys. ${\bf{#1}}$ (#2) #3}

\def\ie{{\it i.e.\/}}
\def\eg{{\it e.g.\/}}

\def\etal{{\it et.al.\/}}

\global\nulldelimiterspace = 0pt

\def\vb#1{{\bf #1}}

\def\L{ {\cal L }}
\def\O{ {\cal O }}

\def\mwd{M_W^2}

\def\mzd{M_Z^2}

\def\mh{m_H}

\def\vtau{\mbox{\boldmath $\tau$}}

\def\ZHr{ \sqrt{Z_H}}

\begin{document}
\pagenumbering{arabic}
\thispagestyle{empty}
\def\thefootnote{\fnsymbol{footnote}}
\setcounter{footnote}{1}

\begin{flushright}
PM/98-34 \\    THES-TP 98/08 \\
short version\\
 \end{flushright}
\vspace{2cm}
\begin{center}
{\Large\bf Tests of Higgs Boson Couplings at a
$\mu^+\mu^-$ Collider}\footnote{Partially supported by the EC contract
CHRX-CT94-0579, and by the NATO grant CRG 971470.}
 \vspace{1.5cm}  \\
{\large G.J. Gounaris$^a$, F.M. Renard$^b$ }\\
\vspace{0.7cm}
$^a$Department of Theoretical Physics, University of Thessaloniki,\\
Gr-54006, Thessaloniki, Greece.\\
\vspace{0.2cm}
$^b$Physique
Math\'{e}matique et Th\'{e}orique,
UMR 5825\\
Universit\'{e} Montpellier II,
 F-34095 Montpellier Cedex 5.\\
\vspace{0.2cm}

\vspace*{1cm}

{\bf Abstract}
\end{center}

We investigate the potential of a muon collider for testing the
presence of anomalous
Higgs boson couplings. We consider the case of a light (less than
$160~GeV$) Higgs boson and  study  the  effects on  the Higgs
branching ratios and  total width, which could be induced by
the non standard couplings created by a class of
$dim=6$ $SU(3)\times SU(2)\times U(1)$ gauge invariant operators
satisfying the constraints  imposed by the present
and  future hadronic and $e^-e^+$ colliders.
For each operator we give the minimal
value of the $\mu^+\mu^-$ integrated luminosity needed for
the  muon collider ($\mu C$) to
improve these constraints. Depending on the operator and
the Higgs mass, this minimal $\mu C$ luminosity lies between
$0.1~fb^{-1}$ and $100~fb^{-1}$.\\


\def\thefootnote{\arabic{footnote}}
\setcounter{footnote}{0}
\clearpage

\section{Introduction}

\hspace{0.7cm}One of the main goals of a future $\mu^+\mu^-$ Collider
($\mu C$), is to
provide a Higgs boson factory \cite{work}.
It has been shown that with a good energy resolution (of the order
of $0.003\%$) and
a reasonable integrated luminosity (of the order of
the $fb^{-1}$), the mass and  the
total width and  branching ratios of a light  Higgs boson
can be directly measured with a high accuracy \cite{Barger}.
In particular, the measurement of the mass and the total width,
is the unique feature of a $\mu C$ \cite{Demarteau}. In addition,
the branching
ratios can be more accurately measured than at a Linear $e^+e^-$
Collider (LC), as long as $\mh \lsim 160~ GeV$;
i.e. as long as the total  width $\Gamma_H$ is sufficiently small,
so that the peak of the $\mu^+\mu^-\to H$ cross section is enhanced,
and a large number of events is produced \cite{Barger,Barger1}.\par

The aim of the present paper is to
discuss more precisely the potential of a
$\mu^+\mu^-$ collider for the search
of anomalous Higgs boson couplings.
We assume that only a single light (\ie\@ $\mh \lsim
160~ GeV$) standard-like Higgs boson exists.
Moreover, we assume that this Higgs boson may have anomalous couplings
which can be generated by adding to the Standard
Model (SM) lagrangian $\L_{SM}$, a set of new physics (NP) terms
associated to a high scale $\Lambda$, lying in the several TeV
range. These NP terms are expressed in terms of all possible
$dim=6$ $SU(3)\times SU(2)\times U(1)$ gauge
invariant operators $\O_i$ involving the various standard model fields
 and contributing with couplings $g^i_{eff}$,
\cite{Hag, Pap, TsiCP}. Thus, the contributions of  each of
these  operators on the partial decay widths $\Gamma(H\to F)$, are
determined by  $g^i_{eff}$.
Constraints on
these coupling constants have already been
established from the effects of the aforementioned operators in the
gauge sector (at LEP1/SLC, LEP2 and TEVATRON)\cite{Hag1,LCWW,WWhad}.
These constraints will be improved by further TEVATRON studies
\cite{Garcia},
as well as studies at
LHC \cite{Hgg} and LC \cite{HZ}
(anticipated to run before  the $\mu C$),
in particular through the direct
production of the Higgs boson.
For each operator we collect
the most stringent constraint on the associated coupling
constant $g^i_{eff}$ that should be available by that time.
We mention separately the constraints that could be obtained
from the study of the $H\gamma\gamma$ couplings, if the
$\gamma\gamma$ mode at a Linear Collider will also be  available
 \cite{LC, LCgg}.\par

Taking the accuracy at which a  $\mu C$
can measure the Higgs total and partial widths,
we subsequently determine the required integrated luminosity $\bar
L(\mu\mu)$ needed in order the $\mu C$ to improved the above
constraints. Hence, for each operator,
we obtain the minimal value of $\bar L(\mu\mu)$ required for this
improvement, as a function of the Higgs mass.\par

The contents of the paper is the following. In Section 2
we list the various $dim=6$ operators affecting the
Higgs couplings and  give the most stringent constraints
expected from
studies at the colliders previous  than $\mu C$.
In Section 3, we describe the effects of these operators
on the Higgs decay widths, and the luminosities required at a $\mu C$,
for improving the previous constraints. For making the
paper self-contained and avoiding normalization uncertainties,
we have collected all the necessary
analytic expressions in Appendix A.
Finally, the results and their implications for the search of
new physics, are summarized in Section 4.\par

\section{The $dim=6$ operators inducing anomalous Higgs couplings}

\hspace{0.7cm}The effective Lagrangian describing
anomalous Higgs properties
is written as:
\bq
     \L_{NP}=\Sigma_i g^i_{eff}\O_i \ \ .
\eq\par

We  first consider purely bosonic operators.
The full list\footnote{We use the linear realization of the scalar
sector, since we are investigating  the case of a light Higgs
particle. }  has been
given in \cite{Hag, Pap, TsiCP}. Retaining  only the
operators affecting the Higgs boson couplings, this list includes
the 8 CP-conserving  operators\footnote{The
contribution of (\ref{listWW}, \ref{listBB}, \ref{listGG})
to the gauge kinetic energy is assumed to have been renormalized
away.}:
\bqa
\O_{\Phi 1} & =& (D_\mu \Phi^\dagger \Phi)( \Phi^\dagger
D^\mu \Phi) \ \ \  , \ \  ~~~g^{\Phi1}_{eff}={f_{\Phi1}\over\Lambda^2}
={\bar{f}_{\Phi1}\over v^2}
 \\[0.1cm]   \label{listPhi1}
\O_{BW} & =& \frac{1}{2}~ \Phi^\dagger B_{\mu \nu}
\overrightarrow \tau \cdot \overrightarrow W^{\mu \nu} \Phi
\ \ \  , \ \ ~~~g^{BW}_{eff}=-{gg'\over2}{f_{BW}\over\Lambda^2}
={\bar{f}_{BW}\over v^2}  \label{listBW}  \\[0.1cm]
\O_{B\Phi} & = & i\, (D_\mu \Phi)^\dagger B^{\mu \nu} (D_\nu
\Phi)\ \ \  , \ \ ~~~ g^{B\Phi}_{eff}={g'f_B\over 2\Lambda^2}=
{g'\bar{f}_B\over 2M^2_W}
={g'\alpha_{B\Phi}\over M^2_W}\label{listBPhi} \\[0.1cm]
\O_{W\Phi} & = & i\, (D_\mu \Phi)^\dagger \overrightarrow \tau
\cdot \overrightarrow W^{\mu \nu} (D_\nu \Phi) \ \ \  ,
\ \ ~~~g^{W\Phi}_{eff}={gf_W\over 2\Lambda^2}={g\bar{f}_W\over 2M^2_W}
={g\alpha_{W\Phi}\over M^2_W}
\label{listWPhi} \\[0.1cm]
\O_{WW} & = &  (\Phi^\dagger \Phi )\,
\overrightarrow W^{\mu\nu} \cdot \overrightarrow W_{\mu\nu} \ \
\  ,  \ \  ~~~g^{WW}_{eff}=-{g^2\over4}{f_{WW}\over\Lambda^2}
={g^2\over4}{d_W\over M^2_W}\label{listWW}\\[0.1cm]
\O_{BB} & = &  (\Phi^\dagger \Phi ) B^{\mu\nu} \
B_{\mu\nu} \ \ \  , \ \ \ ~~~ g^{BB}_{eff}=-{g'^2\over4}{f_{BB}
\over\Lambda^2}
={g^2\over4}{d_B\over M^2_W} \label{listBB} \\[0.1cm]
\O_{\Phi 2} & = & 4 ~ \partial_\mu (\Phi^\dagger \Phi)
\partial^\mu (\Phi^\dagger \Phi ) \ \ \  , \ \
~~~g^{\Phi2}_{eff}={f_{\Phi2}\over8\Lambda^2}
={\bar{f}_{\Phi2}\over v^2}
  \label{listPhi2} \\[0.1cm]
\O_{GG} & = &  (\Phi^\dagger \Phi)\,
\overrightarrow G^{\mu\nu} \cdot \overrightarrow G_{\mu\nu} \ \
\  ,  \ \
~~~g^{GG}_{eff}={d_G\over v^2}    \label{listGG} \ \ ,
\eqa
where
\bq  {g'\over g}={s_W\over c_W} \ \ \ \ , \ \ \ \
 \sqrt{2}G_F\equiv{1\over v^2}\equiv {4M^2_W\over g^2}
\ \ ,
\eq
and  4 CP-violating ones
\bqa
\widetilde{\O}_{BW} & =& \frac{1}{2}~ \Phi^\dagger B_{\mu \nu}
{\vtau} \cdot \widetilde{\vb{W}}^{\mu \nu} \Phi
\ \ ~~~~g^{WB}_{eff}= {\widetilde{\bar{f}}_{BW}\over v^2} \
, \ \  \label{BWtil}\\
\widetilde{\O}_{WW} & = &  (\Phi^\dagger \Phi )\,
\vb{W}^{\mu\nu} \cdot \widetilde{\vb{W}}_{\mu\nu} \ \
~~~~g^{WW}_{eff}={\widetilde{d}_W\over v^2}
\  ,  \ \ \label{WWtil}\\
\widetilde{\O}_{BB} & = &  (\Phi^\dagger \Phi ) B^{\mu\nu} \
\widetilde{B}_{\mu\nu} \ \
~~~~g^{BB}_{eff}={\widetilde{d}_B\over v^2} \
, \ \ \   \label{BBtil} \\
\widetilde{\O}_{GG} & = &  (\Phi^\dagger \Phi )\,
\vb{G}^{\mu\nu} \cdot \widetilde{\vb{G}}_{\mu\nu} \ \
~~~~g^{GG}_{eff}={\widetilde{d}_G\over v^2} \
,  \ \ \label{GGtil}\\[0.1cm]
\eqa
with\footnote{In (\ref{WWtil} -  \ref{GGtil}),
it is understood that instantonic contribution from the
baryon number violating electroweak, as well as the
QCD instantons, have been subtracted, so that only Higgs
interactions are retained.}
$\widetilde V^{\mu\nu}={1\over2}
\epsilon^{\mu\nu\alpha\beta}(\partial_{\alpha}
V_{\beta}-\partial_{\beta}V_{\alpha}) $.\par

We also consider possible modifications of the fermionic couplings
of the Higgs boson. A convenient $dim=6$ operator
for describing anomalous Higgs and heavy quark
interactions was introduced in
\cite{Pap}. In the case of the $b$ quark it reads :\\
\bq
\O_{b1} =  (\Phi^\dagger \Phi )[(\bar{f}_L\Phi)b_R+\bar{b}_R
(\Phi^{\dagger} f_L)]\ \ \  , \ \
~~~g^{b1}_{eff}={ f_{b1}\over\Lambda^2}
={ \bar{f}_{b1}\over v^2} \label{Ob1}
\eq
where $f_L$ is the left handed doublet of the third family of
quarks. The operator $\O_{b1}$  was
motivated by the argument that if NP is associated to the
origin of mass generation, it  should also be characterized by
a  priority in  generating  anomalous couplings  for
the  heavy particles (heavy quarks, possibly heavy leptons
and Higgs bosons) and of course the gauge bosons \cite{Pap}.

In the present work we  also generalize  the operator
$\O_{b1}$,  to  a convenient parametrization of anomalous
$Hff$ couplings for any fermion $f$.
For example for charged leptons we write
\bq
\O_{l1} =  (\Phi^\dagger \Phi )[(\bar{l}_L\Phi)l_R+\bar{l}_R
(\Phi^{\dagger} l_L)]\ \ \  , \ \
~~~g^{l1}_{eff}={ f_{l1}\over\Lambda^2}
={ \bar{f}_{l1}\over v^2} \ \ , \label{Ol1}
\eq
where $l_L,l_R$ are the doublet, singlet of a given family
of lepton.\par

\subsection{\bf Constraints on $g^i_{eff}$ expected to
be established before the Muon Collider run.}

The coupling constants associated to each of these operators are
submitted to constraints obtained or to be
obtained, at present and  future colliders expected to run before
$\mu C$; \ie\@ the LEP, SLC, TEVATRON, LHC, LC in its normal and
$\gamma\gamma$ mode. These  arise from:
\begin{itemize}
\item
virtual bosonic effects in $e^+e^-\to f\bar f$, which  are already
strongly constrained by existing precision measurements at the
Z peak performed by LEP1/SLC \cite{Hag}, and will be further
slightly improved by  measurements at LEP2 and  LC \cite{Hag1}.
\item
direct effects in $e^+e^-\to W^+W^-$  constrained
by LEP2 and  LC measurements \cite{LCWW}; and also
from  $W^+W^-,~ W\gamma,~ WZ$ production at the hadron
colliders  TEVATRON and LHC \cite{WWhad}.
\item
associate Higgs boson production processes
through  $q\bar q'\to HW$ at the TEVATRON \cite{Garcia},
and $e^+e^-\to HZ$
at LEP2 and LC \cite{HZ}.
\item
Higgs production through the process
$gg\to H$ and $gg \to H g$ at LHC \cite{Hgg}.
\item
$\gamma\gamma\to H$, which  should be constrained at an LC
running in the $\gamma\gamma$ \cite{TsiCP}, \cite{GRH}.
\end{itemize}

For each bosonic operator, we have collected in Table 1a,b
the most stringent constraint coming out of the above
list of processes.

\begin{center}
{\bf Table 1a: Upper limits on NP coupling constants: \\
 CP-conserving operators.}\\

\begin{tabular}{|c|c|c|c|c|c|c|c|} \hline
\multicolumn{1}{|c|}{}&
\multicolumn{1}{|c|}{}&
\multicolumn{1}{|c|}{}&
\multicolumn{1}{|c|}{}&
\multicolumn{1}{|c|}{}&
\multicolumn{1}{|c|}{}&
\multicolumn{1}{|c|}{}&
\multicolumn{1}{|c|}{}\\
\multicolumn{1}{|c|}{$|\bar{f}_{\Phi 1}|$}&
\multicolumn{1}{|c|}{$|\bar{f}_{BW}|$} &
\multicolumn{1}{|c|}{$|\bar{f}_{B\Phi}|$} &
\multicolumn{1}{|c|}{$|\bar{f}_{W\Phi}|$} &
\multicolumn{1}{|c|}{$|d_{W}|$}&
\multicolumn{1}{|c|}{$|d_{B}|$}&
\multicolumn{1}{|c|}{$|\bar{f}_{\Phi 2}|$}&
\multicolumn{1}{|c|}{$|d_{GG}|$}
\\ \hline
$0.002$&
$0.0012$&$0.0056$&$0.002$&$0.001$
&$0.0003$&$0.004$&$0.00015$
\\ \hline
\end{tabular}

\end{center}

\vspace{0.5cm}
\begin{center}

{\bf Table 1b: Upper limits on NP coupling constants: \\
 CP-violating operators.}\\

\begin{tabular}{|c|c|c|c|}
\hline
\multicolumn{1}{|c|}{$|\widetilde{\bar{f}}_{BW}|$}&
\multicolumn{1}{|c|}{$|\widetilde{d}_{W}|$} &
\multicolumn{1}{|c|}{$|\widetilde{d}_{B}|$} &
\multicolumn{1}{|c|}{$|\widetilde{d}_{G}|$}
\\ \hline
$0.005$&
$0.004$&$0.0013$&$0.0007$
\\ \hline
\end{tabular}
\end{center}

The fermionic operators in (\ref{Ob1}, \ref{Ol1})
induce purely Higgs-fermion anomalous
interactions. No precise constraint have yet been set on the strength
of this type of couplings.
We  express the coupling strength of these operators
imposing $ f_{f1}=4\pi$  ($f=l,~b$) and defining consequently the
New Physics scale $\Lambda_{NP}$ through
\bq
|g^{f1}_{eff}|={4\pi\over\Lambda_{NP}^2} \ \  . \label{LNP}
\eq
This scale can  be
compared to  other fermionic
scales, like \eg\@  those
obtained from the four-fermion contact interactions in
\cite{contact}.
The measurement of the Higgs branching ratios
at LC should then provide  a lower bound on $\Lambda_{NP}$.
For example, looking at   $e^+e^-\to HZ$ through $B(H\to b\bar b)$
at 250 GeV, with  a luminosity of $\sim 100 fb^{-1}$, for
$m_H=130 GeV$, a lower limit on the scale
$\Lambda_{NP}$ of the order of $40~TeV$ should be possible.
We  take this value as a starting point for
looking at possible improvements with the muon collider.\par

\section{NP effects on the Higgs total width and branching ratios}

\hspace{0.7cm}We now consider the effect of the $dim=6$ operators
on the partial decay widths $\Gamma(H\to F)$.
Each operator is treated separately, and all necessary analytic
expressions are given in the
Appendix. For each interaction term $g^i_{eff}\O_i$,
the \underline{relative} NP effect on
a partial width $\delta^{NP,i}(F)$ is defined through
\bq
\Gamma(H\to F)=\Gamma^{SM}(H\to F)[1+\delta^{NP,i}(F)]
\label{GamNP}
\ \ .
\eq
The magnitude of  $\delta^{NP,i}(F)$
is controlled by the constraints on the coupling constants
$g^i_{eff}$ in Tables 1a,b. The corresponding
 \underline{relative} effect on total Higgs
width $\Gamma_H=\Sigma_{F}\Gamma(H\to F)$ is  given by
\bq
\delta^{NP,i}_H=\Sigma_{F}[B(F)\delta^{NP,i}(F)] \ \ ,
\label{dBH}
\eq
\noindent
while the one on the branching ratios
$B(F)=\Gamma(H\to F)/\Gamma_H$ is
\bq
\delta^{NP,i}_B(F)=\delta^{NP,i}(F)-
\Sigma_{F}[B(F)\delta^{NP,i}(F)] \ \ .
\label{dBF}
\eq
It must be noticed that the NP effect on a given branching ratio
$\delta^{NP,i}_B(F)$, may
come either \underline{directly}
from  the term $\delta^{NP,i}(F)$ in the channel
considered, or \underline{indirectly}
from the NP effect in another channel contributing to the total width;
(\ie\@ to the sum $\Sigma_{F}[B(F)\delta^{NP,i}(F)]$).\par

We next compare these effects with the experimental accuracies
on the various Higgs branching ratios
achievable at a muon collider. Following the
procedure used in \cite{Barger}, we assume a gaussian
$\mu^{\pm}$ beam energy resolution
$\Delta \sim 2~MeV (\sqrt{s}/ 100~GeV)$. Ignoring then initial state
radiation effects,
the peak cross section for the production of
channel $F$ at $\sqrt{s}=m_H$ is given, \cite{Barger2}, by
\bq
\bar{\sigma}(\mu^+\mu^-\to H\to F)\simeq{4\pi\over m^2_H}~
{B(H\to\mu^+\mu^-)B(H\to F)\over[1+{8\Delta^2\over\pi
\Gamma^2_H}]^{1\over2}} \ \ , \label{sigHF}
\eq
\noindent
while  the total cross section is obtained by summing over all final
states $F$ is
\bq
\bar{\sigma}_H \equiv \Sigma_{F}\bar{\sigma}(\mu^+\mu^-\to H\to(F))
\simeq{4\pi\over m^2_H}~
{B(H\to\mu^+\mu^-)\over[1+{8\Delta^2\over\pi
\Gamma^2_H}]^{1\over2}} \label{sigH} \ \ .
\eq

The statistical accuracies at which the measurements of $B(H\to\mu^+\mu^-)$
and $B(H\to F)$ can be achieved, are   computed in terms of the
number of events obtained from the cross sections
of eq.(\ref{sigH}, \ref{sigHF}), and the  integrated luminosity
$\bar{L}(\mu\mu)$. The main channels to
study Higgs decay are the fermionic ones
$f\bar f$ (with $f$ being either a $\mu$ or $\tau$ lepton, or a $c$ or
$b$ quark), and $WW^*$, $ZZ^*$, $Z\gamma$, $\gamma\gamma$ and $gg$.
The  number of events setting the scale of the achievable accuracies
for the various $B(H\to F)$ is evaluated from the SM predictions.
Using the SM parts of the expressions
given in Appendix A, with the QCD corrections defined in
\cite{Spira}, we have reproduced the values of the branching
ratios $B(H\to F)$ obtained in previous works.\par

We   take into account the  background,
due to the $\mu^+\mu^- \to F$
annihilation through processes not involving
a Higgs exchange in the $s$-channel.
 The main background processes (see \cite{Barger}) are
$\mu^+\mu^-\to b\bar b$ (due to $\gamma$ and $Z$ exchange),
$\mu^+\mu^-\to WW^*$ (due to $\nu_{\mu}$, $\gamma$ and $Z$ exchanges),
$\mu^+\mu^-\to ZZ^*$ ($\mu$ exchange), $\mu^+\mu^-\to \gamma\gamma$
($\mu$ exchange) and  $\mu^+\mu^-\to Z\gamma$
(due to $\mu$ exchange). Moreover,  as a background for the
gluon-gluon channel $\mu^+\mu^-\to H \to gg $,
we consider the processes $\mu^+\mu^-\to q\bar q$
($q=u,d,c,s$) due to  $\gamma$ and $Z$ exchanges.

To be reasonably  realistic, we  also
take into account some  detection efficiencies.
These amounts to reducing the number of events  (due to the requirement
of at least one leptonic decay) by the factors:
 $0.33$ for $WW^*$, $0.098$ for
$ZZ^*$, and $0.067$ for $Z\gamma$. This may be somewhat pessimistic,
since  we should keep in mind that some improvement could be obtained
by using hadronic $Z$ modes. For b quarks we use a
detection efficiency of 50\%.  In the $\gamma\gamma$ and $Z\gamma$
channels we  apply an angular cut  of $cos\theta_{cm}<0.7$.\par

Using these , we present  in Tables 2 and 3 below the specific case
of  $m_H=130~GeV$, for which $\Gamma_H\simeq 4.67~MeV$ and
$\bar \sigma_H\simeq 4.\times10^4~fb$, leading through (\ref{sigH})
to a total number of about
$4.\times10^{4}$ Higgs events, for an  integrated
luminosity $\bar L(\mu \mu)$ in the range of $1fb^{-1}$.
The SM branching ratios presented in Table 2, indicate
how these events
are distributed among the various channels, and determine
the achievable accuracies at $\mu C$ indicated in Table 3.
The corresponding accuracies  for the more general case
$0.1 \lsim m_H \lsim 0.18~TeV$, are presented in Fig.1

\begin{center}
{\bf Table 2: SM values of the Higgs branching
ratios for $m_H=130~GeV$}\\[0.5cm]
\begin{tabular}{|c|c|c|c|c|c|c|c|c|c|}
\hline
\multicolumn{1}{|c|}{}&
\multicolumn{1}{|c|}{}&
\multicolumn{1}{|c|}{}&
\multicolumn{1}{|c|}{}&
\multicolumn{1}{|c|}{}&
\multicolumn{1}{|c|}{}&
\multicolumn{1}{|c|}{}&
\multicolumn{1}{|c|}{}&
\multicolumn{1}{|c|}{}&
\multicolumn{1}{|c|}{}
\\[0.001cm]
\multicolumn{1}{|c|}{}&
\multicolumn{1}{|c|}{$\mu^+\mu^-$} &
\multicolumn{1}{|c|}{$\tau^+\tau^-$} &
\multicolumn{1}{|c|}{$b\bar b$} &
\multicolumn{1}{|c|}{$c\bar c$}&
\multicolumn{1}{|c|}{$WW^*$}&
\multicolumn{1}{|c|}{$ZZ^*$}&
\multicolumn{1}{|c|}{$gg$}&
\multicolumn{1}{|c|}{$\gamma\gamma$}&
\multicolumn{1}{|c|}{$Z\gamma$}
\\[0.2cm] \hline
$B(H\to F)$&
$0.00020$&$0.057$&$0.52$&$0.024$&$0.28$&$0.035$&$0.073$&$0.0026$&
$0.0021$\\
[0.1cm] \hline
\end{tabular}
\end{center}

\vspace{0.5cm}

\begin{center}
{\bf Table 3:
Accuracies on $B(H\to F)$ for $m_H=130~GeV$}\\
($\delta_B(F)$ should be multiplied by $\bar{L}(\mu\mu)^{-{1\over2}}$,
 with  $\bar{L}(\mu\mu)$ measured in $fb^{-1}$)\\
\vspace*{0.5cm}
\begin{tabular}{|c|c|c|c|c|c|c|c|c|}
\hline
\multicolumn{1}{|c|}{}&
\multicolumn{1}{|c|}{}&
\multicolumn{1}{|c|}{}&
\multicolumn{1}{|c|}{}&
\multicolumn{1}{|c|}{}&
\multicolumn{1}{|c|}{}&
\multicolumn{1}{|c|}{}&
\multicolumn{1}{|c|}{}
\\[0.001cm]
\multicolumn{1}{|c|}{}&
\multicolumn{1}{|c|}{$\mu^+\mu^-$} &
\multicolumn{1}{|c|}{$b\bar b$} &
\multicolumn{1}{|c|}{$WW^*$} &
\multicolumn{1}{|c|}{$ZZ^*$}&
\multicolumn{1}{|c|}{$gg$}&
\multicolumn{1}{|c|}{$\gamma\gamma$}&
\multicolumn{1}{|c|}{$Z\gamma$}
\\[0.1cm] \hline
$\delta_B(F)$&
$0.015$&$0.020$&$0.022$&$0.084$&$0.085$&$1.2$&$4.8$\\
[0.1cm] \hline
\end{tabular}\\
\end{center}

\vspace{0.5cm}

We next compare the experimental accuracies $\delta_B(F)$
to the corresponding
relative shifts $\delta^{NP,i}_B(F)$
due to NP effects described by the various $dim=6$
operators in Eqs. (\ref{dBF}) and the Appendix, and the results of
Table 1a,b and Fig.1.
Demanding $\delta_B(F)< \delta^{NP,i}_B(F)$ for each operator and
each channel, we obtain the
minimum value of $\bar{L}(\mu\mu)$  required, so that $\mu C$ provides
an improvent of the results of the previous Colliders.
These results are summarized in Fig.2a,b for the CP-conserving bosonic
operators, in Fig.3 for CP-violating
bosonic operators,  and Fig.4 for the fermionic operators.
In all cases only the most efficient channel id indicated.
Finally, in Table 4a,b we repeat these results for the specific case
 $m_H=130~GeV$; (the numbers in parenthesis
refer to the improvements with respect to constraints expected
from measurements in the $\gamma\gamma$ mode of a LC).\par

\begin{center}
{\bf Table 4a: Required $\mu^+\mu^-$ luminosity in $fb^{-1}$
for $m_H=130~GeV$:}\\
{\bf CP-conserving  operators}\\
\vspace*{0.3cm}
\begin{tabular}{|c|c|c|c|c|c|c|c|}
\hline
\multicolumn{1}{|c|}{$\O_{\Phi 1}$}&
\multicolumn{1}{|c|}{$\O_{BW}$} &
\multicolumn{1}{|c|}{$\O_{B\Phi}$} &
\multicolumn{1}{|c|}{$\O_{W\Phi}$} &
\multicolumn{1}{|c|}{$\O_{WW}$}&
\multicolumn{1}{|c|}{$\O_{BB}$}&
\multicolumn{1}{|c|}{$\O_{\Phi 2}$}&
\multicolumn{1}{|c|}{$\O_{GG}$}
\\[0.1cm] \hline
$413$&
$12(27)$&$164$&$53$&$0.7(27)$
&$0.3(32)$&$1.$&$0.8$
\\[0.1cm] \hline
\end{tabular}
\end{center}

\vspace{0.5cm}

\begin{center}

{\bf Table 4b: Required $\mu^+\mu^-$ luminosity in $fb^{-1}$
for $m_H=130~GeV$:}\\
{\bf CP-violating   and fermionic operators}\\
\vspace*{0.3cm}
\begin{tabular}{|c|c|c|c|c|c|}
\hline
\multicolumn{1}{|c|}{$\widetilde{\O}_{BW}$}&
\multicolumn{1}{|c|}{$\widetilde{\O}_{WW}$} &
\multicolumn{1}{|c|}{$\widetilde{\O}_{BB}$} &
\multicolumn{1}{|c|}{$\widetilde{\O}_{GG}$} &
\multicolumn{1}{|c|}{$\O_{b1} (\Lambda_{NP}=50~TeV)$}&
\multicolumn{1}{|c|}{$\O_{\mu1}(\Lambda_{NP}=500~TeV)$}
\\[0.1cm] \hline
$0.2(21)$&
$15(36)$&$2(26)$&$3$&$0.7$
&$1$
\\[0.1cm] \hline
\end{tabular}
\end{center}

\vspace{0.5cm}

A few comments about these results are now in order.
The highest sensitivity arises from the Higgs decay channels
 $\gamma\gamma$, $Z\gamma$ and $gg$,
for which the SM contribution
is depressed by the loop factor $\alpha/\pi$ or $\alpha_s/\pi$.
However, the $Z\gamma$ accuracy
 weakens because we require the use
of the $Z$ leptonic branching ratio. So finally,
the most stringent
constraints arise  for the operators contributing to the
$\gamma\gamma$ and $gg$ channels. Below we comment separately on the
various operators, assuming initially that LC will only work in its normal
$e^-e^+$ mode.\par

$\O_{\Phi 1}$ leads to a wave function renormalization of the Higgs
field which affects all modes, and to a direct $HZZ$ effect.
Only this direct effect on the $ZZ$ channel is accessible
through the study of the branching ratios, but the  accuracy is
reduced due to the small $Z$ leptonic branching ratio.
Therefore, it will be difficult for the muon Collider to
improve the previous  constraints on this operator.

$\O_{BW}$, $\O_{BB}$, $\O_{WW}$ and their CP-violating partners,
 affect directly the $HZZ$, $HZ\gamma$ and
$H\gamma\gamma$ couplings. In these cases the existing
constraints should easily be
improved at $\mu C$. \par

$\O_{B\Phi}$ only affects
the $HZZ$ and $HZ\gamma$ couplings, and it will be more difficult to
improve its constraint.
The high value of the required luminosity could be reduced if
one could use a better efficiency for the $Z\gamma$ channel.
Here, we have pessimistically taken
$\epsilon_{Z\gamma}=0.067$, as
given by the leptonic mode of the $Z$ only. For indication, if
no reduction were applied ($\epsilon_{Z\gamma}=1$), then the required
luminosity would be $11~fb^{-1}$. \par

Since $\O_{W\Phi}$ and
$\O_{WW}$ affect the $HWW$ coupling,  the $WW$ channel will allow to
get better constraints for this operator.
$\O_{GG}$ and $\widetilde{\O}_{GG}$ affect the $Hgg$ mode and
some improvement on the constraints to be set by LHC seems possible.
On the other hand,
$\O_{\Phi,2}$ only leads to a wave function renormalization of the $H$
field, so that  no constraint can be obtained
from branching ratios alone in this case.\par

In the case of the fermionic operators we have expressed the required
luminosity in terms of the scale $\Lambda_{NP}$. For
$\O_{b1}$, using either the  $b\bar b$ or (indirectly)
the  $WW$ channels, a luminosity
of $1~fb^{-1}$ ($10~fb^{-1}$) allows to reach a scale $\Lambda_{NP}$ of
$60~TeV$ ($105~TeV$). For $\O_{\mu1}$, the
sensitivity is much higher, because the SM coupling is reduced
by the small value of the muon mass. Thus,  a luminosity
of $1~fb^{-1}$ ($10~fb^{-1}$) allows to reach a scale $\Lambda_{NP}$ of
$500~TeV$ ($900~TeV$) in this case.
This should allow an important improvement as
compared to the constraints expected from  LC.\par

Finally we note that for $\O_{BW}$, $\O_{WW}$, $\O_{BB}$ and
their CP-violating partners, the results depend also
on whether the $\gamma\gamma$
mode at LC will  run before  the $\mu^+\mu^-$ collider.
In this later case, the required luminosities should lie in the range
of $10-20~fb^{-1}$. Otherwise
a fraction of $fb^{-1}$ would be sufficient.

\subsection{\bf  Additional tests with the total Higgs width. }

We have also  looked at the possible improvements brought
by a measurement of the total Higgs width,
taking a few points around $s=M^2_H$. In \cite{Demarteau} a
relative accuracy of $16\%$ was quoted for a luminosity of $0.4
~fb^{-1}$; while  in \cite{Janot}, a scan with $0.1~fb^{-1}$ should
give an accuracy of about $10~\%$.  We assume that the
relative uncertainty in  the total Higgs width varies
statistically in terms of the number of events and  write
\bq
\bar{\delta}_{\Gamma}= {\delta_{\Gamma}\over\sqrt{\bar{L}(\mu\mu)}}
\ \ ,
\label{dGexp}
\eq
\noindent
where  $\delta_{\Gamma}$ should be of the order of $0.03$.
We then directly compare  $\bar{\delta}_{\Gamma}$, to the relative
NP effect $\delta^{NP,i}_H$ on the total
width defined in eq.(\ref{dBH}).\par

A priori, one could expect an improvement
on the operators
contributing to the main decay modes  ($b\bar b$, $WW$, $ZZ$); as these
channels would  sensibly affect the total width, and
were not much constrained by the study of the branching ratios.
This is the case of
$\O_{\Phi,1}$, $\O_{\Phi,2}$, $\O_{B\Phi}$ and $\O_{b1}$.
However with the accuracy assumed in eq.(\ref{dGexp}), it turns out that
only $\O_{\Phi,2}$ can be constrained,
 (which was not at all constrained by the branching
ratios). Thus, an improvement of the constraint for this operator
 quoted  in Table 1a, will appear as soon as
$\bar{L}(\mu\mu) > 1~fb^{-1}$.\par

For $\O_{\Phi 1}$, an improvement of  the $400~fb^{-1}$
luminosity level required by
the study of the branching ratios, would only appear
if the Higgs is  light
and $\bar{\delta}_{\Gamma}\lsim 0.02/\sqrt{\bar{L}_{\mu\mu}}$.
For $m_H=130~GeV$ and $\bar{L}_{\mu\mu}\simeq100~fb^{-1}$,
this means an accuracy of about $0.01~MeV$ on the
total width, which is probably
impossible to achieve.\par

For $\O_{b1}$ and $\O_{B\Phi}$, an accuracy of about
$\bar{\delta}_{\Gamma} \lsim 0.01/\sqrt{\bar{L}_{\mu\mu}}$ is needed,
in order to improve the results obtained from the study
of the branching ratios. For other
operators, like the CP-violating ones,
an even smaller (rather  unrealistic) $\bar{\delta}_{\Gamma}$
is needed for an improvement
from the total Higgs width measurement to arise.

\section{Conclusions}

\hspace{0.7cm}We have studied under
what conditions a $\mu^+\mu^-$ collider
working as a Higgs factory,
could improve the present and near future constraints on
anomalous Higgs boson couplings.\par

These anomalous couplings are described by
the set of $dim=6$ $SU(3)\times SU(2)\times U(1)$ gauge
invariant operators consisting of the 8 bosonic
CP-conserving ones
$\O_{\Phi 1}$, $\O_{BW}$, $\O_{B\Phi}$, $\O_{W\Phi}$,
$\O_{WW}$, $\O_{BB}$, $\O_{\Phi 2}$, $\O_{GG}$; the 4 bosonic
CP-violating ones $\widetilde{\O}_{BW}$,
$\widetilde{\O}_{WW}$,
$\widetilde{\O}_{BB}$,
$\widetilde{\O}_{GG}$ and the fermionic operators $\O_{b1}$
$\O_{\mu1}$.
For each of these operators we have taken the most stringent
direct or indirect constraints expected  from
studies  at the leptonic  colliders, LEP, SLC,
LC (in its $e^+e^-$ and $\gamma\gamma$
modes) and the hadronic colliders (TEVATRON and LHC), that will run before
the muon collider.\par

We have then looked at the effects of these anomalous couplings on the
Higgs branching ratios and  the total width; and we have
established the minimal integrated luminosity needed for  the $\mu
C$ to improve the constraints on each of the above operators,
imposed by  the Colliders expected to run previously.
This analysis applies to Higgs boson  masses  below the $WW$
threshold ($m_H \lsim 2M_W$), so that $\Gamma_H$ is sufficiently
small and the peak of the cross section sufficiently enhanced,
to make  the rare Higgs decay modes $gg$, $\gamma\gamma$
and $Z\gamma$  observable.\par

In this case one finds (see Fig.2-4) that for most of the operators
(except for $\O_{\Phi,1}$, $\O_{B\Phi}$), an integrated luminosity
 $\bar L_{\mu\mu}\sim {\mbox few} fb^{-1}$ will be sufficient
for improving the previous
constraints.
The special feature of the $\mu^+\mu^-$ collider creating  these
improvements is that it  provides
good accuracies for the modes $\mu^+\mu^-$, $b\bar b$, and $WW^*$.
The first two modes  should allow also to set unique constraints
on the fermionic operators involving the Higgs field.
For example,
a luminosity of $1~fb^{-1}$ would allow to test
fermionic scales of the order of 60, 500
TeV for $\O_{b1}$, $\O_{\mu 1}$. Such scales are higher
than those
accessible at LC, HERA, LHC for the four-fermion operators
\cite{contact}. In addition
these Higgs-fermion operators are of
totally different nature and are perhaps more closely related to
the role of NP in the mass generation mechanism.\par

The accuracy is worse for the $gg$ mode and for the rare modes
$\gamma\gamma$ and $Z\gamma$. Nevertheless,
these channels a very sensitive to
anomalous couplings, because the of the depressed SM contribution,
and  should provide very good constraints on the
related  NP couplings. In fact for most of
the bosonic operators the best
constraints come from these rare modes.\par

Another unique feature of the $\mu^+\mu^-$ collider is to provide
a good measurement of the total Higgs width, that cannot be obtained
by any other means.
This allows to constrain  NP effects leading merely
to a renormalisation of the Higgs couplings, which  cannot be seen
by solely studying the branching ratios; like \eg\@ the effects
of  $\O_{\Phi,2}$.\par

We finally note (using unitarity relations \cite{Pap, unit}),
that the values of the
NP scales to which these new constraints
correspond, lie in the range of several tens of TeV.
This is a domain
where many theoretical models expect NP to show up.

\underline{Acknowledgments}\\
We like to thank A. Blondel and P. Janot for useful
informations about the muon collider project and several
experimental considerations.

\newpage

\underline{Note added in proof}\par
When
establishing the minimal integrated luminosity needed for  the $\mu
C$ to improve the constraints on each of the considered operators,
imposed by the Colliders expected to run previously; we have assumed
that a luminosity of $100~fb^{-1}$ could be accumulated at a linear
$e^+e^-$ collider (LC). After completion of this work we were informed
by H. Schreiber and P. Zerwas that the collider TESLA  considered
at DESY should
be able to accumulate $1~ab^{-1}$ in about 3 years of operation. Such
a performance should allow to improve the determination of the
$HZZ$ coupling and of the Higgs
branching ratios through the process $e^+e^-\to HZ$.
If the photon-photon mode could be put in operation,
a similar improvement on the determination of the $H\gamma\gamma$
couplings will accordingly occur.\par

Such a factor 10 increase in the LC luminosity should reduce by a
factor 3 the upper limits in the coupling
constants
$|\bar{f}_{BW}|$,
$|\bar{f}_{B\Phi}|$,
$|\bar{f}_{W\Phi}|$,
$|d_{W}|$,
$|d_{B}|$,
$|\bar{f}_{\Phi 2}|$,
$|\widetilde{\bar{f}}_{BW}|$,
$|\widetilde{d}_{W}|$,
$|\widetilde{d}_{B}|$ given in Table 1a,b;
which in turn means that the minimum $\mu C$ luminosity required
for the operators
 $\O_{BW}$, $\O_{B\Phi}$, $\O_{W\Phi}$,
$\O_{WW}$, $\O_{BB}$, $\O_{\Phi 2}$, $\widetilde{\O}_{BW}$,
$\widetilde{\O}_{WW}$,
$\widetilde{\O}_{BB}$ should also be increased by a factor of 10.
For example the values lying between 0.1 and 10
$fb^{-1}$ for these operators in Fig.1,2 would now lie between
1 and 100 $fb^{-1}$.\par

The characteristic  features of the $\mu^+\mu^-$ collider in
providing
a very good measurement of the total Higgs width, and of the
$\mu^+\mu^-$ branching ratio
and the Higgs partial widths,
remain unchanged in this new comparison.

\newpage
\renewcommand{\theequation}{A.\arabic{equation}}
\renewcommand{\thesection}{A.\arabic{section}}
\setcounter{equation}{0}
\setcounter{section}{0}

{\large \bf Appendix A: The partial  decay
widths of the Higgs boson}

\vspace{0.5cm}

In this appendix we define the partial widths for the decay
$H\to F$ of an off-shell Higgs particle  by
\bq
\Gamma_{H\to F}(s) =\frac{(2\pi)^4}{2\mh}
\int |T_{H\to F}(s)|^2 d\Phi_F \ ,
\eq
where $\Phi_F$ gives the usual definition of the invariant phase
space \cite{PDG}. Note that the off-schellness only appears in
the invariant amplitude $T_{H\to F}(s)$.

\section{ $H\to\gamma \gamma$.}

Contributions to this process arise from the  SM at
1 loop \cite{Higgshunter}, and from operators $\O_{BW}$,
$\O_{WW}$, $\O_{BB}$, their CP-violating partners, and also
$\O_{\Phi,1}$,  $\O_{\Phi,2}$ from $Z_H$. The result is
\bqa
 \Gamma_{H\to\gamma \gamma}(s) &=& {\sqrt2 G_F \over 16\pi m_H} s^2
\Big\{  \Big |{\alpha\over4\pi}({4\over3}F_t + F_W)\ZHr
 - 2d_W s^2_W-2d_B c^2_W
+\bar{f}_{BW} s_W c_W\Big |^2 \nonumber\\
&&+ \Big |2\widetilde{d}_W s^2_W+2\widetilde{d}_B c^2_W
-\widetilde{\bar{f}}_{BW} s_W c_W\Big |^2 \Big\} \ , \ \ \ \ \
\eqa
in which the Higgs wave function renormalization $Z_H$ is
determined by  the tree level NP contribution of $\O_{\Phi,1}$,
$\O_{\Phi,2}$ and is given by
\bq
Z_H=[1+8\bar{f}_{\Phi,2}+{1\over2}\bar{f}_{\Phi,1}]^{-1} \ \ ,
\label{RZH}
\eq
\noindent
while the standard contribution arises from top and $W$
loops  respectively determined by
\bq
F_t = -2t_t(1+(1-t_t)f(t_t))   \ \ \ \ \ , \ \ \ \ \
\eq
\bq
F_W = 2+3t_W+3t_W(2-t_W)f(t_W) \ \ \ \ \ \ , \ \ \ \
\eq
in terms of
\bqa
f(t) = \left [\sin^{-1}(1/\sqrt{t})\right]^2 \ \ \ \ \ \ \ \
\ \ &
\makebox{\ \ \ \ \ if \ \ \ } &
           t \geq 1 \ \ \ \ , \ \ \ \nonumber \\[0.5cm]
f(t)= -{1\over4}\left [\ln \left ({1+\sqrt{1-t} \over 1-\sqrt{1-t}}
\right )
-i\pi\right ]^2 & \makebox{\ \ \ \ if \ \ \ } & t < 1 \ \
\ \ , \ \ \ \  \label{gamgam}
\eqa
where  $t_t =4m^2_t/s$ and  $t_W=4m^2_W/s$.

\section{ $H\to\gamma Z$.}

Contributions arise here also  from the SM
1 loop top and $W$ contributions \cite{Higgshunter},
as well as from  $\O_{BW}$,
$\O_{WW}$, $\O_{BB}$, their CP violating analogs, and also
from the operators $\O_{B\Phi}$ ,$\O_{W\Phi}$,   $\O_{\Phi,1}$,
$\O_{\Phi,2}$. The result is
\bqa
\Gamma_{H\to \gamma Z}(s)  &=& {\sqrt2G_F s^2 \over 8 \pi m_H}
\Big (1-{\mzd
\over s }\Big )^3\Bigg (  \Big |{\alpha\over
4\pi}(A_t+A_W)\ZHr +2(d_W-d_B)s_Wc_W\nonumber \\
&&-{1\over2}(c^2_W-s^2_W)\bar{f}_{BW}
- {s_W\over2c_W}(\bar{f}_B-\bar{f}_W)\Big |^2
\nonumber \\
&& + \Big |2(\widetilde{d}_W-\widetilde{d}_B)s_Wc_W
-{1\over2}(c^2_W-s^2_W)\widetilde{\bar{f}}_{BW}\Big |^2\Big )
\ ,
\eqa
with
\bq
A_t =
{(-6+16s^2_W)\over 3s_Wc_W}[I_1(t_t,l_t)-I_2(t_t,l_t)]
\ \ \ \  , \ \ \ \ \ \
\eq
\bq
A_W
=-\cot\theta_W[4(3-\tan^2\theta_W)I_2(t_W,l_W)+[(1+{2\over
t_W})
\tan^2\theta_W-(5+{2\over t_W})]I_1(t_W,l_W)] \ ,
\eq
where $t_t =4m^2_t/s$, $t_W=4\mwd/s$ as before, and
$l_t=4m^2_t/ \mzd$ , $l_W=4\mwd/ \mzd$, and
\bq
I_1(a,b)={ab\over 2(a-b)}+{a^2b^2\over
2(a-b)^2}[f(a)-f(b)]+{a^2b\over(a-b)^2}[g(a)-g(b)] \  ,
\eq
\bq
   I_2(a,b)=-{ab\over2(a-b)}[f(a)-f(b)]\ \ \ \ , \
\eq
 $f(t)$ is given in (\ref{gamgam}) and
\bqa
\ \ \ \ \ g(t)=\sqrt{t-1} \sin^{-1}({1\over\sqrt{t}}) &
\makebox{ \ \ \ if \ \ \ }  &
t\geq 1 \ \ \ , \ \nonumber \\[.2cm]
g(t) ={1\over2}\sqrt{1-t}\left[ \ln\left({1+\sqrt{1-t}
\over1-\sqrt{1-t}}\right)-i\pi\right] & \makebox{\ \ \ if \ \ \ }
& t< 1 \ \ \ \ \ . \ \ \
\eqa \par

\section{ $H\to gg$.}

Contributions arise from the SM 1-loop top exchanges
and from tree level contribution of
the operators  $\O_{GG}$ and $\O_{\Phi1}$,
$\O_{\Phi2}$. The result is

\bq
\Gamma_{H\to gg}(s)= {s^2
\over8\pi m_H}[1+({95\over4}-{7N_F\over6}){\alpha_s\over\pi}]
[ |A_{SM}\ZHr -{4d_G\over v}|^2+|{4\widetilde{d}_G\over v}|^2]
\ , \
\eq
where
\bq
A_{SM}= -{ \alpha_s t_t \over 2\pi v }
 (1+(1-t_t) f(t_t)) \ ,
\eq
with   $t_t= 4m_t^2/s$ and $f(t)$  given in (\ref{gamgam}).
Note the presence of an  important QCD correction factor
which, (for  the number of light quark flavours $N_F=5$) is
of the order of 65\%.

\section{ $H\to WW$.}

Contributions arise from the SM at tree level
and from operators
$\O_{W\Phi}$, $\O_{WW}$, $\widetilde{\O}_{WW}$,as well as from
 $\O_{\Phi1}$, $\O_{\Phi2 }$ which induce a  wave function
renormalization of the Higgs field.  For $m_H>2M_W$, we get
\bqa
\Gamma_{H\to W^+W^-}(s) & = &
 {\alpha  \beta_W \over 16s^2_W M^2_W m_H}
\Big (2 \left [2M^2_W\ZHr -2d_W(s-2 M^2_W) - \bar{f}_W s
\right ]^2\nonumber \\
&&+ \left [ \ZHr (s-2 M^2_W)-4d_W M^2_W - \bar{f}_W s
\right ]^2\nonumber \\
&&+8|\widetilde{d}_W|^2 s(s-4M^2_W)\Big )
\eqa
\noindent
in which $\beta_W=\sqrt{1-4M^2_W/s}$.

For $M_W<m_H<2M_W$, the  Higgs decay width is computed with
one virtual gauge boson decaying into a lepton or
quark pair. The expression is \cite{Higgshunter, HVV*, Hgg}
\bqa
\Gamma_{H \to WW^*}(s)~&=&~\frac{3\alpha^2 s}{32 \pi m_H s^4_W}~
[(\sqrt{Z_H}-{\bar{f}_W\over2x})^2 D_{SM}(x) +d_W\sqrt{Z_H}D_1(x)
-\bar{f}_W\sqrt{Z_H}D_4(x) \nonumber\\
&&+8d_W^2D_2(x)+\bar{f}^2_W D_5(x) -d_W\bar{f}_W D_6(x)
+8|\widetilde{d}_W|^2D_3(x)]
\eqa
\noindent
where $x={m^2_W/s}$ and
\bqa
D_{SM}(x) &= &\frac{3(20x^2-8x+1)}{\sqrt{4x-1}}\cos^{-1}
\left(\frac{3x-1}{2x^{3/2}}\right) \nonumber \\
& \null &
  - ~(1-x)\left(\frac{47x}{2} -
\frac{13}{2} +\frac{1}{x} \right) -
3(2x^2-3x+\frac{1}{2})\ln(x) \ \ \ \ \
\ , \ \eqa\\
\bqa
D_1(x) &=&\frac{24(14x^2-8x+1)}{\sqrt{4x-1}}\cos^{-1}
\left(\frac{3x-1}{2x^{3/2}}\right) \nonumber \\
 & \null & + ~12(x-1)(9x-5)-12(2x^2-6x+1)\ln(x) \ \ \ \ \ , \ \ \
\eqa\\
\bqa
D_2(x) &= &\frac{54x^3-40x^2+11x-1}{x\sqrt{4x-1}}\cos^{-1}
\left(\frac{3x-1}{2x^{3/2}}\right) \nonumber \\
& \null &
+~ \frac{(x-1)}{6}(89x-82+\frac{17}{x})-(3x^2-15x+\frac{9}{2}-
\frac{1}{2x})\ln(x)
\ \ \ \ \ ,
\eqa
\bqa
D_3(x) &= &\frac{-28x^2+11x-1}{x\sqrt{4x-1}}\cos^{-1}
\left(\frac{3x-1}{2x^{3/2}}\right) \nonumber \\
& \null &
  - ~\frac{x^2}{6} -
\frac{21x}{2} +\frac{27}{2} -\frac{17}{6x} +
{(6x^2-9x+1)\over 2x}\ln(x) \ \ \ \ \
\ , \
\eqa
\bqa
D_4(x) &= &-~ {\sqrt{4x-1}}~\frac{(10x-4)}{x}\cos^{-1}
\left(\frac{3x-1}{2x^{3/2}}\right) \nonumber \\
& \null &
-~ \frac{(x-1)}{3x^2}(2x^3+50x^2-31x+3)+(6x-9+\frac{2}{x})\ln(x)
\ \ \ \ \
\ , \
\eqa
\bqa
D_5(x) &= & \frac{3 \sqrt{4x-1}}{4x^2 }\cos^{-1}
\left(\frac{3x-1}{2x^{3/2}}\right) \nonumber \\
& \null &
+{1\over16x^3}(1-x)(x^4-7x^3-9x^2-25x+4)-{3\over4x^2}(x^2+x-
{1\over2})\ln(x)  \ \ \ \ , \
\eqa
\bqa
D_6(x) &= &{4(14x^2-13x+2)\over x\sqrt{4x-1}} \cos^{-1}
\left(\frac{3x-1}{2x^{3/2}}\right) +{2(1-x)\over3x}(x^2-17x+28)
\nonumber \\
& \null &
+2(9-{2\over x})\ln(x)
\ \ \ \ \ \ . \
\eqa \par

\section{ $H\to ZZ$.}

Contributions arise from the SM at tree level
and from operators $\O_{BW}$, $\O_{B\Phi}$
$\O_{W\Phi}$, $\O_{WW}$, $\O_{BB}$, $\widetilde{\O}_{BW}$,
$\widetilde{\O}_{WW}$, $\widetilde{\O}_{BB}$ as well as
 $\O_{\Phi1}$, $\O_{\Phi2 }$ through the wave function
renormalization of the Higgs field.
For $m_H>2M_Z$, we get
\bqa
\Gamma_{H\to ZZ}(s) & = &
 {\alpha  \beta_Z \over 32s^2_W M^2_W m_H}
\Bigg (2 \Big [2M^2_Z(\sqrt{Z_H}+\bar{f}_{\Phi1})
\nonumber \\
&&-(2d_B s^2_W+2d_W c^2_W+\bar{f}_{BW}s_Wc_W)(s-2
M^2_Z)
-(\bar{f}_W +\bar{f}_B{s^2_W\over c^2_W})s \Big ]^2
\nonumber \\
&&+\Big [(\sqrt{Z_H}+\bar{f}_{\Phi1})(s-2 M^2_Z)-
2M^2_Z(2d_B s^2_W+2d_W
c^2_W  \nonumber \\
 && +\bar{f}_{BW}s_Wc_W)
-(\bar{f}_W +\bar{f}_B{s^2_W\over c^2_W})s
\Big ]^2\nonumber \\
&&+2\left |(2\widetilde{d}_B s^2_W+2\widetilde{d}_W c^2_W
+\widetilde{\bar{f}}_{BW}s_Wc_W)\right |^2M^2_H(s-4M^2_Z)
\Bigg ) \ ,
\eqa
\noindent
with $\sqrt{Z_H}$ given by eq.(\ref{RZH})
and $\beta_Z=\sqrt{1-4M^2_Z/s}$.\\

For $M_Z<m_H<2M_Z$, we get
\bqa
\Gamma_{H \to ZZ^*}(s)&=& \frac{\alpha^2 s}{128 \pi m_H s^4_Wc^4_W}~
\left (7-{40s^2_W\over3}+{160s^4_W\over9}\right )
\cdot  \nonumber \\
&&
\Bigg [\left (\sqrt{Z_H}+\bar{f}_{\Phi,1}
-{1\over2x}(\bar{f}_W+\bar{f}_B{s^2_W\over c^2_W})
\right )^2D_{SM}(x) \nonumber\\
&& +(\sqrt{Z_H}+\bar{f}_{\Phi1})
\left (d_Bs^2_W+d_W c^2_W+{\bar{f}_{BW}\over2}c_Ws_W\right )D_1(x)
\nonumber \\
&&-(\sqrt{Z_H}+\bar{f}_{\Phi,1})\left (\bar{f}_W
+\bar{f}_B{s^2_W\over c^2_W} \right )D_4(x)
\nonumber \\
&& +2(2d_Bs^2_W+2d_W c^2_W+\bar{f}_{BW}s_Wc_W)^2D_2(x)
+\left (\bar{f}_W+\bar{f}_B{s^2_W\over c^2_W}
\right)^2 D_5(x)\nonumber\\
&& +{1\over2}(2d_Bs^2_W+2d _Wc^2_W+\bar{f}_{BW}s_Wc_W)
\left (\bar{f}_W+\bar{f}_B{s^2_W\over c^2_W}\right )D_6(x)\nonumber\\
&&+2|(2\widetilde{d}_B s^2_W+2\widetilde{d}_W c^2_W
+\widetilde{\bar{f}}_{BW}s_Wc_W)|^2D_3(x)\Bigg ]
\eqa
\noindent
where $x=M^2_Z/s$.

\section{ $H\to f\bar f$.}

Contributions arise from the SM at tree level,
from operators
$\O_{\Phi,1}$, $\O_{\Phi,2 }$ through the wave function
renormalization of the Higgs field $Z_H$, (compare (\ref{RZH})),
and from the fermionic operator $\O_{f1}$. We find
\bq
\Gamma_{H\to f\bar f}(s)={N_c s\over8\pi m_H}\beta_f^3~
\left |-{m_f\over v}
\sqrt{Z_H}+{3\over2\sqrt{2}}\bar{f}_{f1}\right |^2
\eq
with the colour factor
\bqa
N_c & = &1 \ \ \ \ \   \ \ \ \ \ \ \ \ \
\ \ \ \mbox{  for leptons } \ ,\\
N_c& =& 3(1+5.67{\alpha_s\over\pi})\ \ \ \   \mbox{ for quarks} \ ,
\eqa
and $\beta_f = \sqrt{1-{4m^2_f/s}}$. In the case of quarks ($f=q$),
the mass $m_f$ is the running mass $m_q(m_H)$ computed with the
expression given in ref.\cite{Spira}.

\newpage

\begin{figure}[p]
\vspace*{-3cm}
\[
\epsfig{file=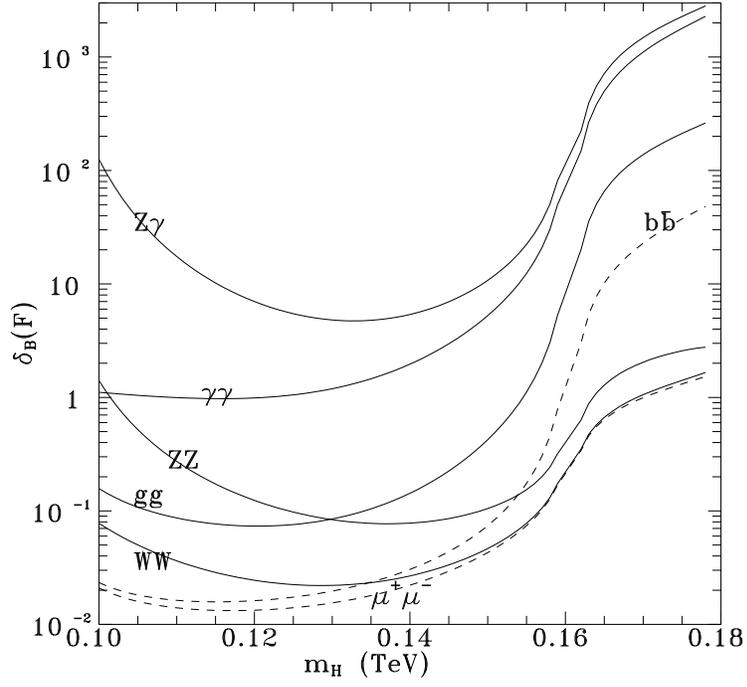,height=9cm}
\]
\vspace*{1.cm}
\caption[1]{Accuracies on the various branching ratios
$B(H\to F)$ versus the Higgs mass.
(The indicated values
should be multiplied by $\bar{L}(\mu\mu)^{-{1\over2}}$,
the integrated luminosity $\bar{L}(\mu\mu)$ being
measured in $fb^{-1}$).}
\label{Figure1}
\end{figure}

\newpage
\begin{figure}[p]
\vspace*{-3cm}
\[
\epsfig{file=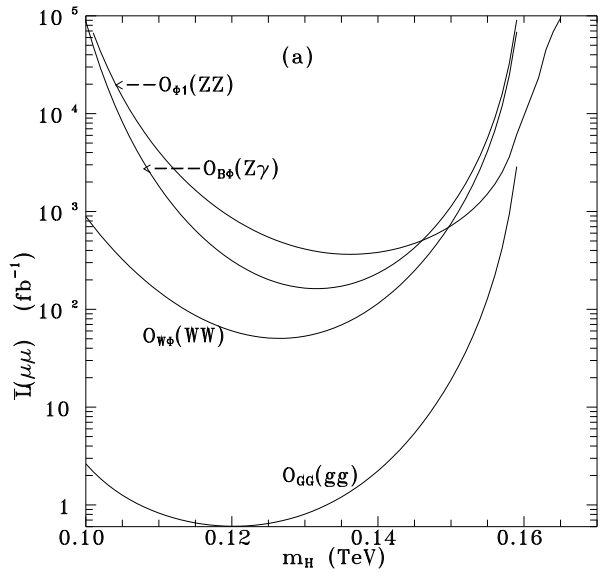,height=7.5cm}
\]
\vspace*{1cm}
\[
\epsfig{file=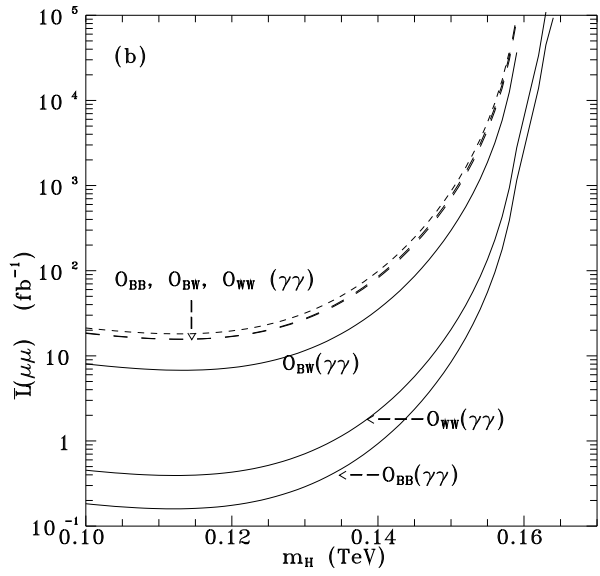,height=7.5cm}
\]
\vspace*{1.cm}
\caption[1]{$\mu^+\mu^-$ luminosity needed to improve the
constraints on the CP-conserving operators obtained from an LC used in
the $e^+e^-$ (solid) or Laser backscattering (dash) mode. Only the most
efficient decay channel is indicated.}
\label{Figure2ab}
\end{figure}

\newpage
\begin{figure}[p]
\vspace*{-3cm}
\[
\epsfig{file=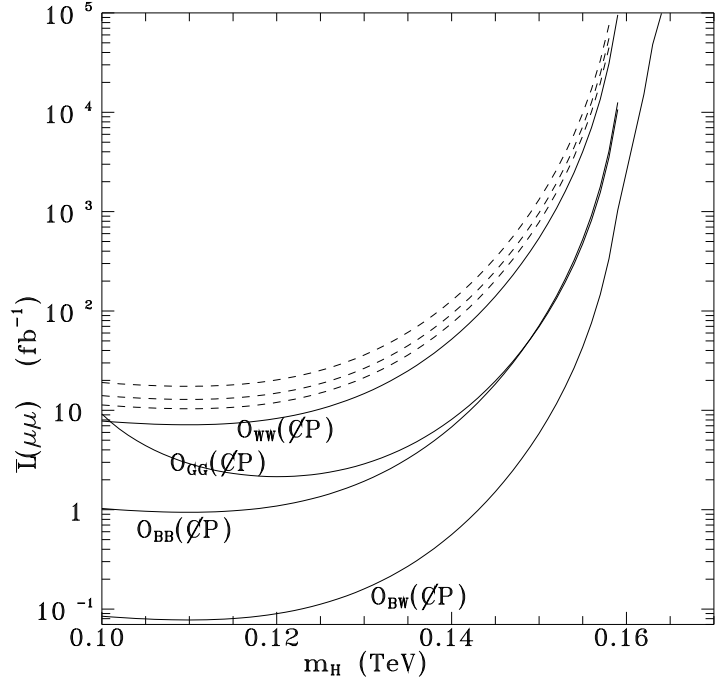,height=9cm}
\]
\vspace*{1.cm}
\caption[1]{$\mu^+\mu^-$ luminosity needed to improve the
constraints on the CP-violating  operators obtained from an LC used in
the $e^+e^-$ (solid) or Laser backscattering (dash) mode. Only the $\gamma
\gamma $ decay channels is indicated, which is the most efficient.
The dash line contributions of are for $\O_{BB}$, $\O_{BW}$ and
$\O_{WW}$ and have the same relative ordering as the solid-line results.}
\label{Figure3}
\end{figure}

\clearpage
\newpage

\begin{figure}[p]
\vspace*{-3cm}
\[
\epsfig{file=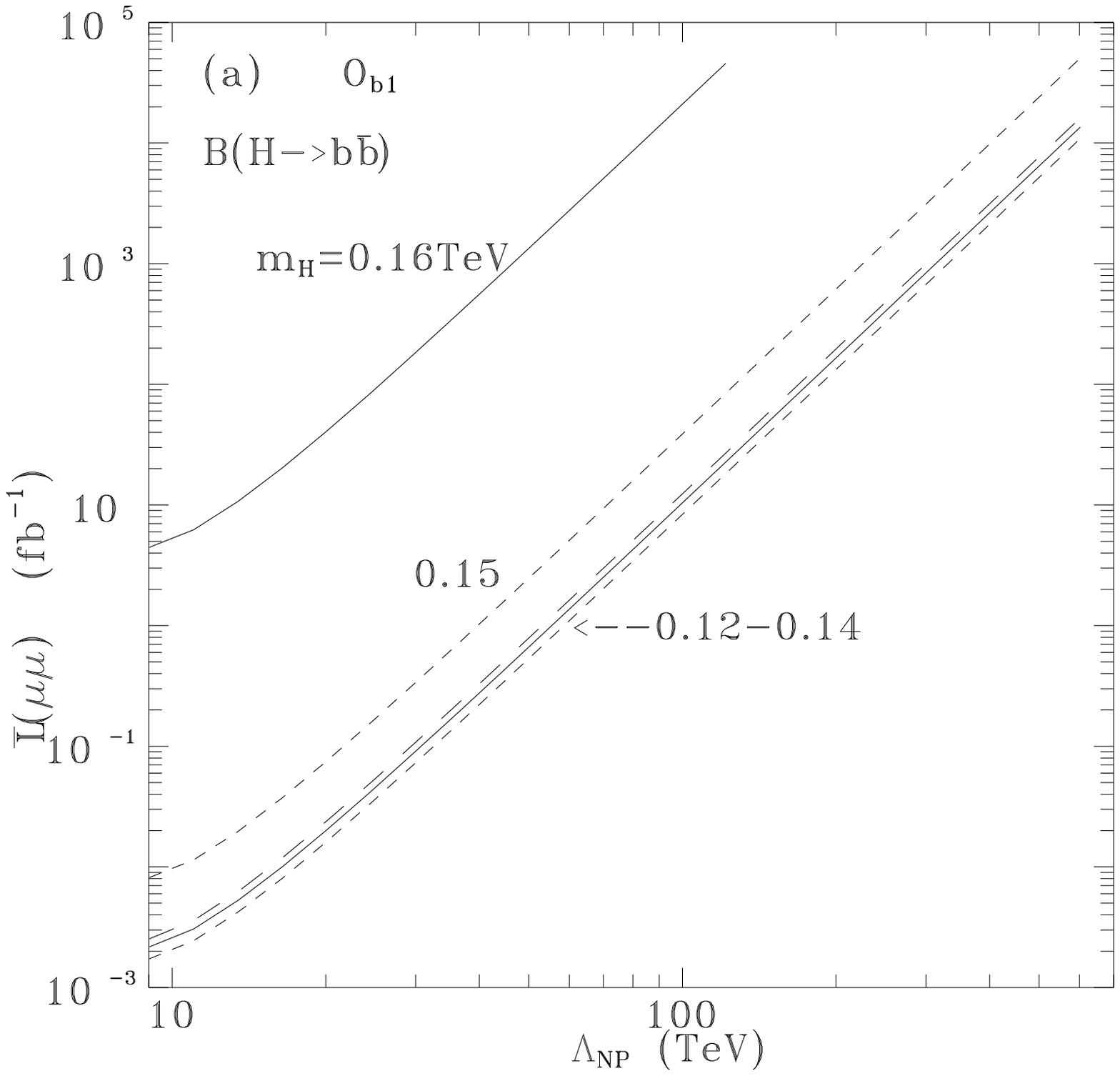,height=7.5cm}
\]
\vspace*{1cm}
\[
\epsfig{file=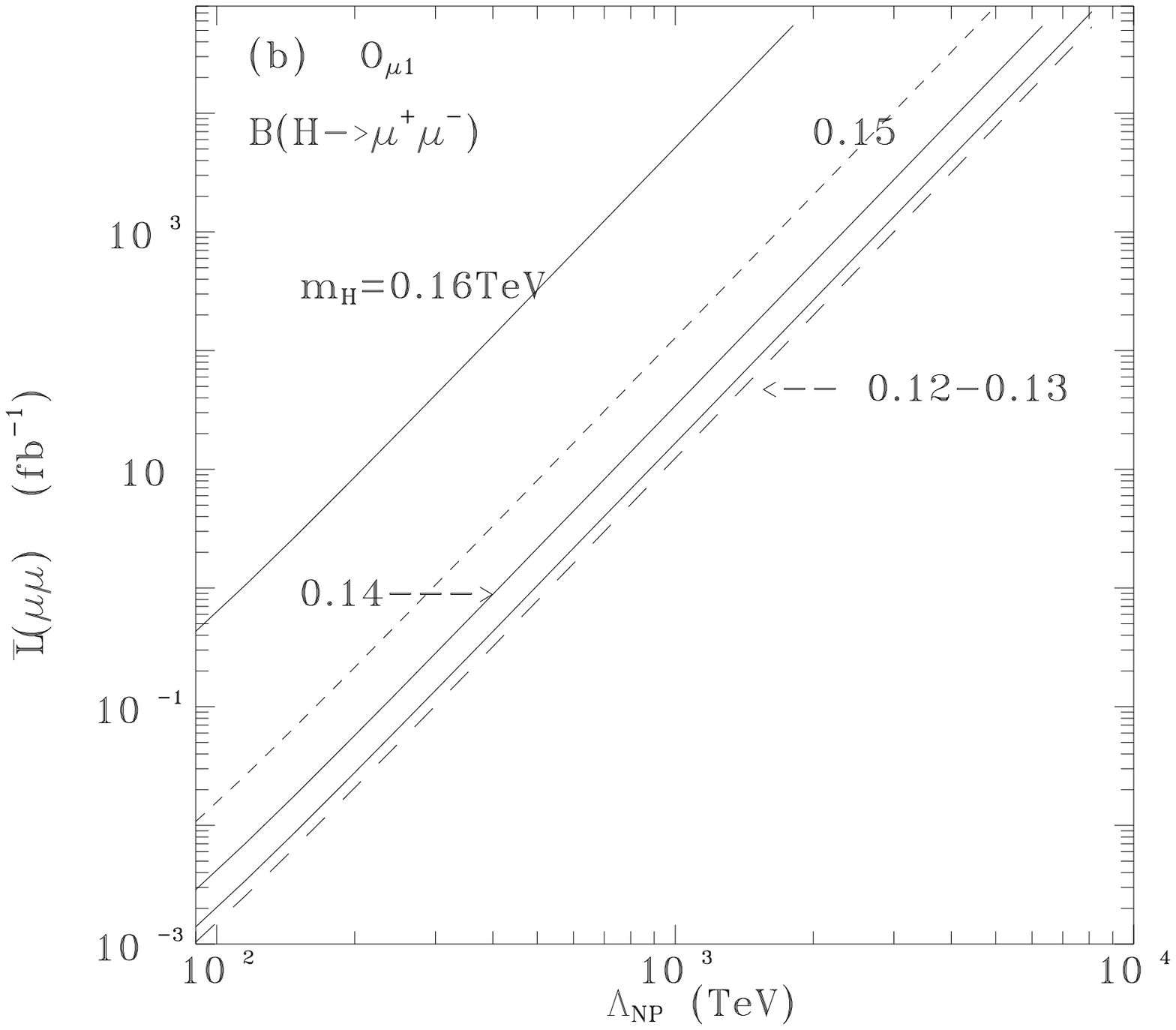,height=7.5cm}
\]
\vspace*{1.cm}

\caption[1]{$\mu^+\mu^-$ luminosity needed to reach the new
physics scale $\Lambda_{NP}$ corresponding to the
operator $\O_{b1}$ (a), $\O_{\mu1}$ (b), for the indicated decay
channel.}
\label{Figure4}

\end{figure}

\end{document}